\documentclass[aps,prl,twocolumn,superscriptaddress,showpacs]{revtex4-1}

\usepackage{dcolumn} % Align table columns on decimal point
\usepackage{graphicx}% Include figure files
\usepackage{dcolumn} % Align table columns on decimal point
\usepackage{bm}      % bold math
\usepackage{color}
\usepackage{amsmath,amssymb}
\usepackage{multirow}
\usepackage{epstopdf}
\usepackage{latexsym}
\usepackage{subfigure}
\usepackage{dsfont} %for ensemble alphabet, and works with numbers too, use \mathds{1}
\usepackage{wasysym} %for hexagon (in Eq environment or in text)
\usepackage{times}
\usepackage{epstopdf}
\usepackage{gensymb}
\usepackage{textcomp}
\usepackage{lipsum}

\usepackage[draft]{changes}

\begin{document}
\title{Ba$_8$CoNb$_6$O$_{24}$: a spin-1/2 triangular-lattice Heisenberg antiferromagnet in the 2D limit}

\author{R.~Rawl}
\affiliation{Department of Physics and Astronomy, University of Tennessee, Knoxville, Tennessee 37996, USA}

\author{L.~Ge}
\affiliation{School of Physics, Georgia Institute of Technology, Atlanta, GA 30332, USA}

\author{H.~Agrawal}
\affiliation{Quantum Condensed Matter Division, Oak Ridge National Laboratory, Oak Ridge, Tennessee 37381, USA}

\author{Y.~Kamiya}
\affiliation{Condensed Matter Theory Laboratory, RIKEN, Wako, Saitama 351-0198, Japan}

\author{C.~R.~Dela Cruz}
\affiliation{Quantum Condensed Matter Division, Oak Ridge National Laboratory, Oak Ridge, Tennessee 37381, USA}

\author{N.~P.~Butch}
\affiliation{NIST Centre for Neutron Research, National Institute of Standards and Technology, Gaithersburg, MD 20899, USA}

\author{X.~F.~Sun}
\affiliation{Hefei National Laboratory for Physical Sciences at Microscale, University of Science and Technology of China, Hefei, Anhui 230026, People's Republic of China}
\affiliation{Key Laboratory of Strongly-Coupled Quantum Matter Physics, Chinese Academy of Sciences, Hefei, Anhui 230026, People's Republic of China}
\affiliation{Collaborative Innovation Center of Advanced Microstructures, Nanjing, Jiangsu 210093, People's Republic of China}

\author{M.~Lee}
\affiliation{Department of Physics, Florida State University, Tallahassee, FL 32306, USA}
\affiliation{National High Magnetic Field Laboratory, Florida State University, Tallahassee, FL 32310, USA}

\author{E.~S.~Choi}
\affiliation{National High Magnetic Field Laboratory, Florida State University, Tallahassee, FL 32310, USA}

\author{J.~Oitmaa}
\affiliation{School of Physics, The University of New South Wales, Sydney, NSW 2052, Australia}

\author{C.~D.~Batista}
\affiliation{Department of Physics and Astronomy, University of Tennessee, Knoxville, Tennessee 37996, USA}
\affiliation{Quantum Condensed Matter Division and Shull-Wollan Center, Oak Ridge National Laboratory, Oak Ridge, Tennessee 37831, USA}

\author{M. Mourigal}
\email{mourigal@gatech.edu}
\affiliation{School of Physics, Georgia Institute of Technology, Atlanta, GA 30332, USA}

\author{H.~D.~Zhou}
\email{hzhou10@utk.edu}
\affiliation{Department of Physics and Astronomy, University of Tennessee, Knoxville, Tennessee 37996, USA}
\affiliation{National High Magnetic Field Laboratory, Florida State University, Tallahassee, FL 32310, USA}

\author{J. Ma}
\email{jma3@sjtu.edu.cn}
\affiliation{Department of Physics and Astronomy, Shanghai Jiao Tong University, Shanghai 200240, China}
\affiliation{Department of Physics and Astronomy, University of Tennessee, Knoxville, Tennessee 37996, USA}

\date{ \today}

\begin{abstract}
The perovskite Ba$_8$CoNb$_6$O$_{24}$ comprises equilateral effective spin-1/2 Co$^{2+}$ triangular layers separated by six non-magnetic layers. Susceptibility, specific heat and neutron scattering measurements combined with high-temperature series expansions and spin-wave calculations confirm that Ba$_8$CoNb$_6$O$_{24}$ is basically a two-dimensional (2D) magnet with no detectable spin anisotropy and no long-range magnetic ordering down to 0.06 K. In other words,  Ba$_8$CoNb$_6$O$_{24}$ is very close to be a realization of the paradigmatic spin-$1/2$ triangular Heisenberg model, which is not expected to  exhibit symmetry breaking at finite temperature according to the Mermin and Wagner theorem.
\end{abstract}
\maketitle
	
\setlength{\parskip}{0em}

In a celebrated 1966 paper~\cite{Mermin_1966}, Mermin and Wagner demonstrated that thermal fluctuations prevent 2D magnets to spontaneously break their continuous spin-rotation symmetry if the  interactions decay fast enough with the distance between spins. The role of thermal fluctuations is replaced by quantum fluctuations in one-dimensional (1D) systems at  temperature $T=0$; for instance, the spin-1/2  Heisenberg antiferromagnetic chain does not display long-range magnetic order in the $T=0$  limit and instead hosts quasi-long-range correlations~\cite{Lieb_1961} and fractional spin excitations~\cite{Faddeev_1981,Tennant_1993,Mourigal_2013}. Quantum fluctuations are also expected to have a strong effect on the ground states of highly frustrated 2D and 3D Mott insulators. Indeed, the realization of {\it quantum spin-liquids}, quantum-entangled states of matter which do not exhibit magnetic ordering, is a major focus of modern condensed matter physics~\cite{Savary_2016,Banerjee_2016}. While spin-liquids are an extreme case of quantum states of matter, 2D systems that  \textit{do} order at $T=0$ can still exhibit strong deviations from semi-classical behavior. For instance, the elementary excitations of a 2D ordered magnet (magnons) become  weakly bonded pairs  of fractional excitations near the ``quantum melting point" (QMP) that signals the transition into a spin liquid state. A clear indication of  proximity to a QMP is a strong suppression of the ordered moment relative to the full moment.

The spin-1/2 2D triangular-lattice Heisenberg antiferromagnet (QTLHAF) displays non-collinear spin-order at $T\!=\!0$ with a relative suppression of the ordered moment of more than 50\% ~\cite{Jolicoeur89,Chubukov94,Capriotti99,Zheng06,White06,Chernyshev_2009}. This makes it an ideal model for studying the effect of strong quantum fluctuations on  the spectrum of magnetic excitations. In real materials, however, weak interlayer interactions and spin or spatial anisotropies are likely present. Even extremely small perturbations are sufficient to induce long-range magnetic order at a sizable  N\'eel temperature $T_{\rm N}$, because $T_{\rm N}$ increases logarithmically in the interlayer-coupling or in the exchange anisotropy~\cite{r,w,b,m,o}. This is the case for well-studied compounds comprising transition-metal ions, such as Cs$_2$CuCl$_4$ ($T_{\rm N}\!=\!0.62$~K~\cite{Coldea_1997}) and Ba$_3$CoSb$_2$O$_9$  ($T_{\rm N}\!=\!3.8$~K~\cite{2d}). Quantum effects remain prominent below $T_{\rm N}$ and lead to order from disorder phenomena, such as the one third magnetization plateaux~\cite{2h,1i,Koutroulakis2015},  in the presence of an external magnetic field.  A recent inelastic neutron scattering (NS)  study of Ba$_3$CoSb$_2$O$_9$~\cite{2m} showed that even in presence of sizable perturbations~\cite{2c,2e,2d,2i} relative to the pure QTLHAF, dynamical features are not captured by spin-wave theory (SWT). This observation suggests that alternative theoretical approaches  are not only needed to describe spin-liquid states, but also to  account for qualitative properties of the excitation spectrum of ordered magnets near their QMP~\cite{2o,Ghioldi_2015}.

% =========================
\begin{figure*}[th!]
	\linespread{1}
	\par
	\begin{center}
		\includegraphics[width= 5.5 in]{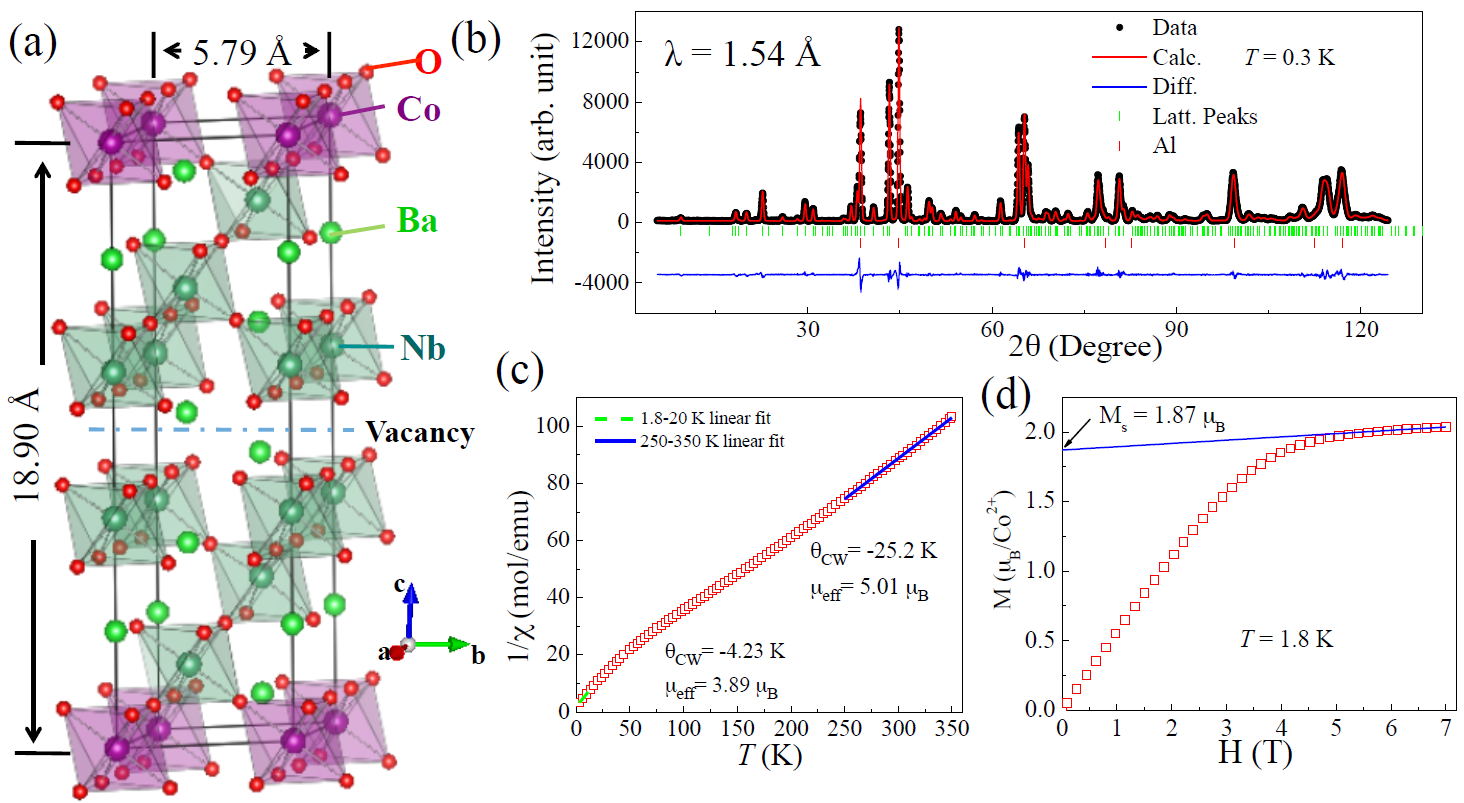}
	\end{center}
	\par
	\caption{\label{diff}%
      (color online)
      (a) Stacked layer structure of Ba$_8$CoNb$_6$O$_{24}$ with Co$^{2+}$ ions sitting on a triangular lattice. (b) Rietveld refinement of the neutron powder diffraction pattern measured at $T$ = 0.3 K with $\lambda$ = 1.54 {\AA}. (c) Temperature dependence of the inverse DC magnetic susceptibility and corresponding Curie-Weiss fits. (d) Isothermal DC magnetization measured at $T$ = 1.8 K and extrapolation of the saturated magnetization from the linear dependence above the saturation field (blue solid line).}
\end{figure*}
% =========================

In this Letter, we introduce Ba$_8$CoNb$_6$O$_{24}$, a new realization of the QTLHAF  model obtained from Ba$_3$CoSb$_2$O$_9$ by intercalating non-magnetic layers between the triangular planes. We present structural, thermo-magnetic, inelastic NS and theoretical results indicating that spin-space anisotropy and inter-plane interactions are both essentially absent in Ba$_8$CoNb$_6$O$_{24}$. Having a model realization of the QTLHAF at hand, we test predictions from semi-classical spin-wave theory and investigate potential exotic phenomena arising from enhanced quantum fluctuations.

To obtain Ba$_8$CoNb$_6$O$_{24}$, we start from Ba$_3$CoSb$_2$O$_9$, a compound that comprises layers of magnetic CoO$_6$ octahedral stacked along the hexagonal $c$ axis and  separated by two non-magnetic SbO$_6$ layers. The intra-layer Co--Co distance is 5.86~{\AA} and the interlayer Co--Co distance is 7.23~{\AA} \cite{2d}. In the former material, the  inter-layer  magnetic exchange interaction, $J^{\prime}$, is $\sim$5\% of the  intra-layer exchange $J$~\cite{2e,2m}.  Moreover, Ba$_3$CoSb$_2$O$_9$ possesses a small easy-plane XXZ anisotropy (the ratio between the longitudinal and transverse exchange interactions is $\Delta\!\approx\!0.9$). While the degree of spin anisotropy is difficult to control, one natural strategy to reduce the  inter-layer interaction is to insert additional non-magnetic layers in between the magnetic layers. Ba$_8$CoNb$_6$O$_{24}$ exactly meets these requirements: it contains a vacant layer and \emph{six layers} of non-magnetic NbO$_6$ octahedral between triangular layers of Co$^{2+}$ ions  [see Fig. 1(a)]. While the intra-layer Co--Co distance of 5.79~{\AA}  is comparable to Ba$_3$CoSb$_2$O$_9$, the inter-layer Co--Co distance is dramatically increased up to 18.90~{\AA}~\cite{2p}. This  remarkable structure  is expected to guarantee a true 2D nature for the magnetic properties of Ba$_8$CoNb$_6$O$_{24}$.

To confirm the physical outcome of our intercalation strategy, we  present structural and thermo-magnetic characterization of powder samples of Ba$_8$CoNb$_6$O$_{24}$ grown from a solid-state synthesis method detailed in the supplemental information (SI)~\cite{supp}. A fit to our neutron powder diffraction (NPD) pattern measured at $T\!=\!0.3$~K with $\lambda\!=\!1.54$~{\AA} [Fig. 1(b)] yields the space-group $P\bar{3}m1$ with {$a$ = 5.7902(2) {\AA} and $c$ = 18.9026(3) {\AA}}. A Rietveld refinement yields structural parameters given in SI \cite{supp} and indicates a limited amount of disorder ($<\!2$~\%) between the Co and Nb sites, consistent with  an earlier study \cite{2p}. The patterns at $T\!=\!0.3$~K and 2.0 K are essentially identical: no additional Bragg peaks appear and broadening of existing peaks is not observed within the sensitivity and resolution of our experiment~\cite{supp}, suggesting  the  absence of a structural transition or long-range magnetic order down to  $T\!=\!0.3$~K.

The temperature dependence of the magnetic DC susceptibility, $\chi(T)$, shows no sign of magnetic ordering or spin freezing down to $T\!=\!1.8$~K [Fig.~1(c)]. The slope of $1/\chi(T)$ changes around $T\!=\!150$~K; Curie-Weiss fits yield $\mu_{\rm eff}\!=\!5.01(2)$~$\mu_{\rm B}$ and $\theta_{\rm CW}\!=\!-25.2(3)$~K for 200 K $\!<\!T\!<\!$ 350 K, and $\mu_{\rm eff}\!=\!3.89(2)$~$\mu_{\rm B}$ and $\theta_{\rm CW}\!=\!-4.23(1)$~K for 1.8 K $\!<\!T\!<\!$ 30 K. The effective moment reduction indicates a crossover from a high-spin state ($S\!=\!3/2$) to a low-spin state ($S\!=\!1/2$) and is typical for Co$^{2+}$ ions in an octahedral environment, see \textit{e.g.}, ACoB$_3$ (A = Cs, Rb, B = Cl, Br)~\cite{ACoB3}. The isothermal DC magnetization at $T\!=\!1.8$~K, shown in Fig.~1(d), indicates that spins saturate above $\mu_0 H_s\!\approx\!4$~T, while a fit to the linear magnetization observed from $\mu_0 H = 5$~T to 7 T uncovers a Van Vleck paramagnetic contribution of 0.023 $\mu_{\rm B}$.T$^{-1}$ per Co$^{2+}$ and yields a saturation magnetization $M_{\text{s}} = 1.87$~$\mu_B$.  This value is comparable to that of Ba$_3$CoSb$_2$O$_9$ and corresponds to a powder-averaged gyromagnetic ratio $g$ = 3.84 for the effective $S\!=\!1/2$ Kramers doublet.
% =========================
\begin{figure}[h!]
	\linespread{1}
	\par
	\begin{center}
		\includegraphics[width=3.0 in]{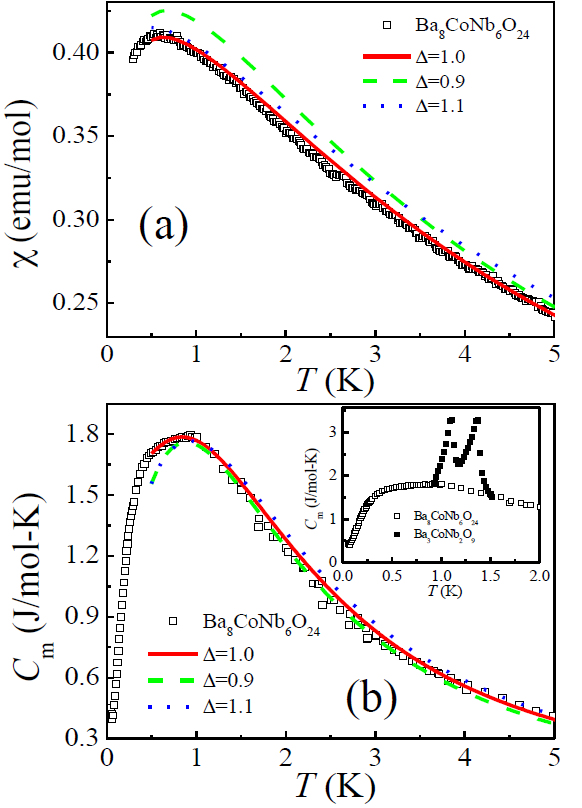}
	\end{center}
	\caption{\label{diff}%
      (color online)
      (a) Temperature dependence of the  magnetic AC susceptibility of Ba$_8$CoNb$_6$O$_{24}$ and  corresponding high-temperature series expansion simulations for the 2D spin-1/2 triangular-lattice antiferromagnet with XXZ exchange anisotropy. Values of $\Delta$ = 0.9, 1.0, and 1.1 are used and simulations run down to a temperature of 0.5 K using Pad\'e approximants of order [6,6].  The measurements are obtained with an AC excitation field of amplitude 0.5 Oe and frequency 300\,Hz, and matched to the DC susceptibility below $T\!=\!15$~K by an overall $T$-independent rescaling factor~\cite{supp}.  (b) Temperature dependence of the magnetic part of the specific heat of Ba$_8$CoNb$_6$O$_{24}$ and matching simulations. (Insert) Comparison to the magnetic specific heat of Ba$_3$CoNb$_2$O$_9$.}
\end{figure}
% =========================

 Similarly, the $T$-dependence of the magnetic AC susceptibility, shown in Fig.~2(a), uncovers no sharp features down to $T\!=\!0.3$~K. Instead, it reveals a broad peak centered at $T\!=\!0.6$~K, which we associate with the onset of short-range magnetic correlations. The presence of  magnetic correlations below $T\!\approx\!1$~K is confirmed by the heat-capacity measurements shown in Fig.~2(b). The magnetic contribution to the specific heat, $C_{\text{m}}$, was isolated by subtracting the lattice contribution, $C_{\text{L}}$, of the iso-structural non-magnetic compound Ba$_8$ZnTa$_6$O$_{24}$ \cite{supp}. The  $C_{\text{m}}(T)$ curve reveals a broad peak around $T\!=\!0.8$~K without any sharp feature down to $T\!=\!0.06$~K (the small increase at lower temperatures is attributed to nuclear spins), suggesting the absence of a magnetic phase transition  down to $T\leq0.06$~K. By integrating $C_m(T)/T$ from $T_{\textrm{min}}=0.06$~K to a target  ($ T\leq8$~K), we obtain the change in magnetic entropy $\Delta S_m = S_m(T) - S_m(T_\textrm{min})$~\cite{supp}. The release of entropy reaches 5.32 J\,mol$^{-1}$\,K$^{-1}$ at $T=8$~K, which is close to the value $R\ln{2}$ = 5.76 J\,mol$^{-1}$\,K$^{-1}$ expected for a Kramers doublet ground-state.

What is the origin of the broad peak observed in $C_{\text{m}}(T)$? Previous quantum Monte Carlo studies on quasi-2D antiferromagnetic Heisenberg models have shown that the onset of long-range magnetic order yields a sharp peak in $C_{\text{m}}(T)$ even for inter-layer exchange interactions as small as $J^{\prime}/J$ = 2$\times$10$^{-4}$~\cite{2g}. Upon further decreasing the inter-layer coupling, the sharp peak disappears and only a broad peak remains. This is precisely the behavior we observe in Ba$_8$CoNb$_6$O$_{24}$, thus exposing the practically ideal 2D nature of magnetism in this compound. This becomes even clearer when our results are compared to Ba$_3$CoNb$_2$O$_9$ [see the inset of Fig.~2(b)], which comprises  only two non-magnetic layers between the magnetic planes. The specific heat of  Ba$_3$CoNb$_2$O$_9$ reveals two subsequent phase transitions at $T_{\text{N1}}$ = 1.10 K and $T_{\text{N2}}$ = 1.36 K, indicating  the presence of easy-axis anisotropy \cite{2q}. At a similar energy scale ($\approx$ 1 K), Ba$_8$CoNb$_6$O$_{24}$ only exhibits a single broad peak with no observable signs of exchange anisotropy or inter-layer coupling.

 The temperature dependence of $\chi(T)$ and $C_{\text{m}}(T)$ for the  QTLHAF model has been well documented using high-temperature series expansions (HTSE)~\cite{2t,htse1,htse2,htse3}  up to 12th order \cite{2a}.  To determine if exchange anisotropy is present in Ba$_8$CoNb$_6$O$_{24}$, we extend existing HTSE work to the XXZ Hamiltonian,
\begin{equation}
	\displaystyle{
	\label{eq:H}
	\mathcal{H}=J\sum_{\langle i,j \rangle} (S_i^xS_j^x+S_i^yS_j^y+\Delta\, S_i^zS_j^z)},
\end{equation}
where $\langle i,j \rangle$ denotes nearest-neighbor  spins. We obtained results for the isotropic ($\Delta\!=\!1.0$), easy-plane ($\Delta\!=\!0.9$), and easy-axis ($\Delta\!=\!1.1$) models~\cite{supp}. The best HTSE fit to our experimental observations, namely $\chi(T)$ and $C_{\rm m}(T)$ below $T=$ 5 K, yields $J=0.144$~meV for $\Delta\!=\!1.0$ with a fitting error-bar on $J$ smaller than $10^{-3}$~meV [see Fig.~2]. For a fixed value of $J$, the fit quality becomes worse as soon as $\Delta$ deviates from 1.0 and lead to higher (respectively lower) peak heights for $\chi(T)$ (resp. $C_{\text{m}}$).

% =========================
\begin{figure*}[t!]
	\linespread{1}
	\par
	\begin{center}
		\includegraphics[width=5.9 in]{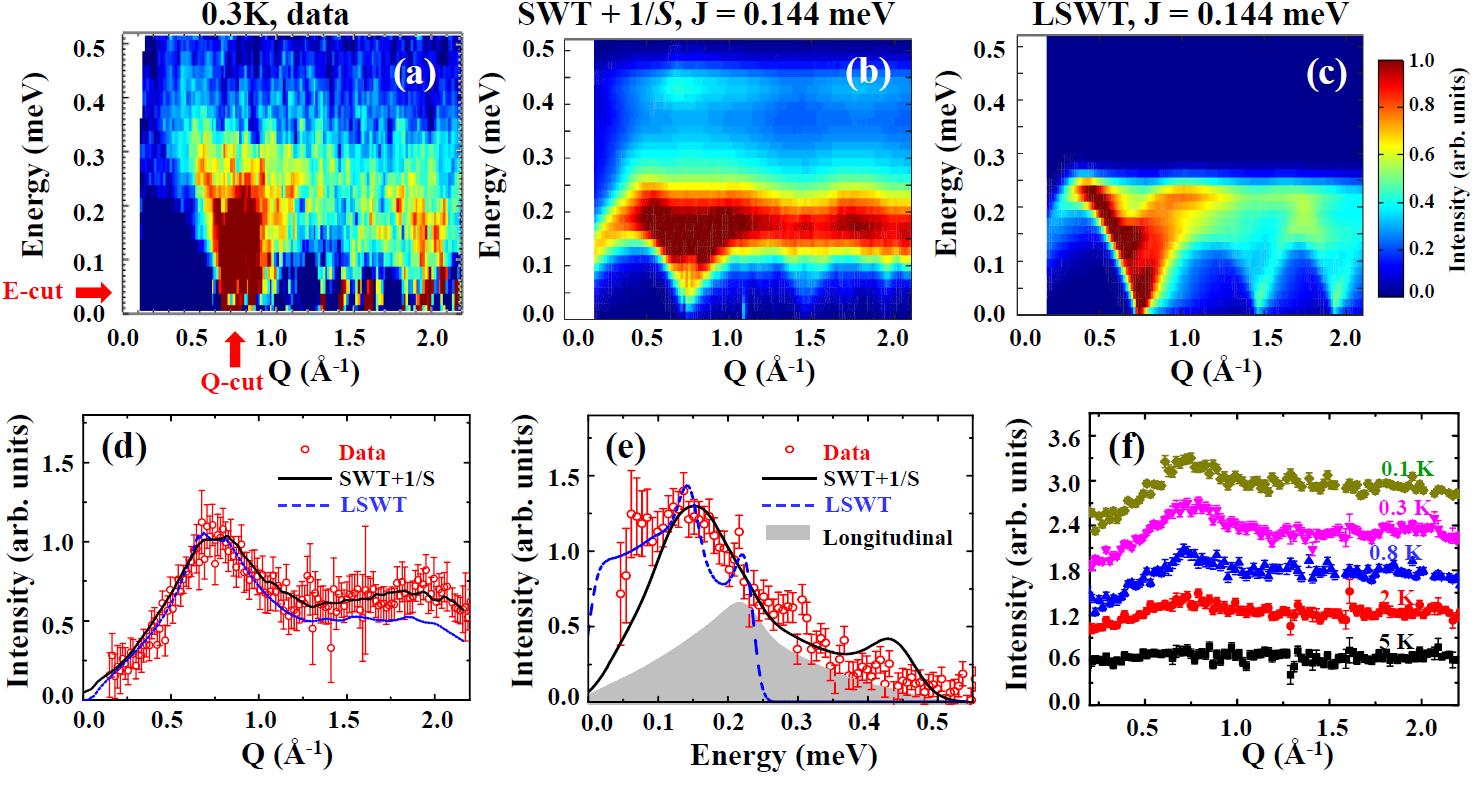}
	\end{center}
	\caption{\label{diff}%
      (color online)
      (a) Powder-averaged inelastic NS spectra of Ba$_8$CoNb$_6$O$_{24}$ at $T = 0.3\,\text{K}$. Data collected at $T = 10\,\text{K}$ is used as background. (b,c) NS intensity calculated for $J=\text{0.144\,meV}$ using non-linear SWT with 1/$S$-corrections and linear SWT, respectively. Calculated intensities have been convoluted by Gaussian profiles of full-width at half maximum $\Delta E\!=\!0.025$ meV and $\Delta Q\!=\!0.015$ \AA$^{-1}$ to approximate the effects of instrumental resolution. (d,e) Comparisons between experiment (red dots), 1/$S$-SWT (solid black line) and linear SWT (dashed blue line) as energy-integrated ($0.05\leq\!E\!\leq0.52$~meV) and momentum-integrated ($0.6\leq\!Q\!\leq0.9$ \AA$^{-1}$) cuts, respectively. The shaded (gray) area corresponds to the longitudinal (two-magnon) contribution to the NS intensity in 1/$S$-SWT. The high-energy bump around $E=$0.45~meV in (e) is an artifact of our 1/$S$ approximation~\cite{tafdsf}. (f) Temperature dependence of the energy-integrated intensity of (d). Error bars correspond to one standard error.}
 \end{figure*}
 % =========================

With strong  thermodynamic indication that Ba$_8$CoNb$_6$O$_{24}$ realizes the purely 2D and spin-isotropic QTLHAF model, we now turn to the nature of its spin excitations. NS intensity ( powder-averaged) as a function of momentum transfer $Q$ and energy transfer $E$ allows to track the development of magnetic correlations upon lowering $T$. In Fig.~3(a), we present such results for $T = 0.3\,\text{K}$,  with additional results for $5$~K $\leq T\leq 0.05$~K included in SI~\cite{supp}. The momentum dependence of the magnetic signal reveals strong ridges of intensity emerging from $Q\!\approx\!0.7$ \AA$^{-1}$ with less intense repetitions at $1.5$ \AA$^{-1}$ and $2.0$ \AA$^{-1}$. While spins appear well-correlated at $T\!=\!0.3$~K, the low-energy signal ($E\leq0.1$~meV) remains broader than instrumental resolution suggesting that spin correlations remain short-ranged and static magnetic order is absent. The energy dependence of the main signal reveals  gapless  excitations extending up to 0.35 meV with less intense signal reaching up to $E=0.45$~meV.  These features do not change significantly as $T$ is lowered~\cite{supp}.

To model the dynamic magnetic correlations, we resort to SWT at $T\!=\!0$; $1/S$ corrections~\cite{tafdsf} are included in Fig.~3(b) while we remain strictly at the linear level (LSWT)~\cite{spinw} in Fig.~3(c). We assume that the system orders in the 120$^\circ$ magnetic structure, at least at $T=0$, and use $J\!=\!0.144$~meV from the thermodynamic measurements. Our $E$-integrated [Fig.~3(d)] and $Q$-integrated [Fig.~3(e)] scans reveal a good agreement between NS measurements and powder-averaged $1/S$-SWT predictions. The most visible improvement between $1/S$ and linear SWT calculations stems from the inclusion of longitudinal spin fluctuations in the former. These excitations reflect the reduction of the ordered moment by quantum fluctuations and form a high-energy continuum also known as two-magnon scattering. The absence of notable temperature dependence for the $E\!\geq\!0.1$~meV magnetic scattering below $T\!=\!0.5$~K [Fig.~3(f)] further supports the evidence for strong quantum fluctuations in the ground-state of Ba$_8$CoNb$_6$O$_{24}$.

It is instructive to compare  the excitations of Ba$_8$CoNb$_6$O$_{24}$ with  that of the quasi-2D compound Ba$_3$CoSb$_2$O$_{9}$, for which $J^\prime\!=\!0.05~J$, $J\!\approx\!1.7$\,meV, and $\Delta\!\approx\!0.9$. While both compounds comprise structurally similar magnetic layers with comparable Co--Co bond lengths, the $\sim\!2.0$~meV in-plane excitation bandwidth of Ba$_3$CoSb$_2$O$_{9}$ is an order of magnitude larger than the present observation of $\sim\!0.18$ meV for Ba$_8$CoNb$_6$O$_{24}$. In units of their respective $J$, the bandwidth $W\!\approx\!1.18~J$ for the former compound compares well with $W\!\approx\!1.24~J$ obtained by the present $1/S$-SWT analysis for Ba$_8$CoNb$_6$O$_{24}$ [see Fig.~3(b)]. While Ba$_3$CoSb$_2$O$_{9}$ develops long-range magnetic ordering below $T_{\rm N}=3.7$~K $\sim 0.19J$, Ba$_8$CoNb$_6$O$_{24}$ does not exhibit any magnetic ordering  down to $T=0.06$~K $\sim 0.04J$. Given that $T_{\rm N}$ increases logarithmically both in the magnitude of $J^\prime$ and $\Delta$, the suppression of $T_{\rm N}/J$ by a factor of at least 4 relative to Ba$_3$CoSb$_2$O$_{9}$ implies that inter-plane and anisotropic exchange interactions must be extremely small in Ba$_8$CoNb$_6$O$_{24}$.

In conclusion, our powder-sample experiments reveal that Ba$_8$CoNb$_6$O$_{24}$ is virtually an ideal realization of the QTLHAF and an unique compound to expose the consequences of the Mermin and Wagner theorem in a real triangular-lattice material. Recent studies have shown that quantum fluctuations have a non-perturbative effect on the magnetic excitations of quasi-2D quantum antiferromagnets~\cite{2m,DallaPiazza}. We expect even stronger quantum effects in the magnetic excitation spectrum of Ba$_8$CoNb$_6$O$_{24}$, making it an even better candidate to challenge existing semi-classical theories for the dynamic response of frustrated quantum antiferromagnets. From the materials discovery standpoint, our work devises a method for reducing dimensionality by intercalating non-magnetic layers in layered compounds that can be extended to other lattices to reveal new physics.

\begin{acknowledgments}
 R.R. and H.D.Z. thank the support of  NSF-DMR-1350002. J.M. thanks the support of the Ministry of Science and Technology of China (2016YFA0300500). The work at Georgia Tech (L.G., M.M.) was supported by the College of Sciences and ORAU's Ralph E. Powe Junior Faculty Enhancement Award. X.F.S. acknowledges support from the National Natural Science Foundation of China (Grant Nos. 11374277 and U1532147), the National Basic Research Program of China (Grant Nos. 2015CB921201 and 2016YFA0300103), and the Opening Project of Wuhan National High Magnetic Field Center (Grant No. 2015KF21).The work at NHMFL is supported by NSF-DMR-1157490, the State of Florida and the U.S. Department of Energy. The work at ORNL High Flux Isotope Reactor was sponsored by the Scientific User Facilities Division, Office of Basic Energy Sciences, U.S. Department of Energy. \vspace{-0.7cm}
\end{acknowledgments}

% -------------------------------------------------------------------
% Supplementary materials
% -------------------------------------------------------------------
\clearpage
\onecolumngrid
\appendix
%\doublespace
\setcounter{figure}{0}
\setcounter{table}{0}
\setcounter{equation}{0}
\renewcommand{\thefigure}{S\arabic{figure}}
\renewcommand{\thetable}{S\arabic{table}}
\renewcommand{\theequation}{S\arabic{equation}}

% -------------------------------------------------------------------

\begin{center}
{\large \bf Supplementary online material for  ``Ba$_8$CoNb$_6$O$_{24}$: a spin-1/2 triangular-lattice Heisenberg antiferromagnet in the 2D limit''}
\end{center}

\textit{Experimental details---}%
Ba$_8$CoNb$_6$O$_{24}$ was synthesized using solid state reactions by mixing stoichiometric ratios of BaCO$_3$, CoCO$_3$, and Nb$_2$O$_5$ and annealing at 1500 $^\circ$C for 48 hours with one intermediate grinding.  The DC susceptibility measurements were taken down to 1.8 K using a commercial superconducting interference device magnetometer (Quantum Design) in a field of 5000 Oe. The AC susceptibility was measured at National High Magnetic Field Laboratory using a homemade setup with AC excitation field 0.5 Oe and frequency 300 Hz \cite{AC}. The specific heat was measured down to 0.05 K using a Physical Properties Measurement System of Quantum Design. The neutron powder diffraction (NPD) was carried out down to 0.3 K on HB-2A neutron powder diffractometer at Oak Ridge National Laboratory with wavelengths of 1.5406 {\AA} and 2.4111 {\AA} with a collimation of 12$^\prime$-open-6$^\prime$. The shorter wavelength gives a greater intensity and higher Q coverage that was used to investigate the crystal structure, while the longer wavelength gives lower Q coverage and greater resolution that was important for investigating the magnetic structures. Rietveld refinement was performed using Fullprof to refine the data \cite{Fprof}. Inelastic neutron scattering (INS) was performed down to 0.06 K at National Institute of Standards and Technology (NIST) CHRNS using the time-of-flight spectrometer, disk chopper spectrometer (DCS) with three wavelengths of 1.8, 5.0, and 7.5 {\AA} \cite{Copley}. \\

\textit{Neutron powder diffraction---}%
Fig. S1(a) shows the neutron powder diffraction patterns measured at 0.3 and 2 K with  $\lambda$ =  2.4111 {\AA}. There is no resolvable differences between them and thus no signs for long range magnetic ordering (LRO) or structural distortion down to 0.3 K. The lattice parameters and atomic positions refined from 0.3 K data are listed in Table S1.

\begin{table*}[tbph!]
\par
\caption{Structural parameters for Ba$_8$CoNb$_6$O$_{24}$ at 0.3 K}
\label{t1}
\setlength{\tabcolsep}{0.25cm}
\begin{tabular}{ccccccc}
\hline
\hline
Refinement & Atom & Site & {\it x} & {\it y} & {\it z} & Occupancy \\ \hline
\multirow{5}{*}{\begin{tabular}[c]{@{}c@{}}$\lambda$ = 1.5404 {\AA}\\ RF-factor = 2.44\\ Bragg R-fct = 3.21\\ $P$-$3m1$\end{tabular}}
& Ba(1) & 2c & 0 & 0 & 0.18771(36) & 1/6 \\
& Ba(2) & 2d & 1/3 & 2/3 & 0.06126(52) & 1/6 \\
& Ba(3) & 2d & 1/3 & 2/3 & 0.45478(41) & 1/6 \\
& Ba(4) & 2d & 1/3 & 2/3 & 0.68149(32) & 1/6 \\
& Co & 1a & 0 & 0 & 0 & 0.084(2) \\
& Nb(1) & 2c & 0 & 0 & 0.38677(30) & 0.166(1) \\
& Nb(2) & 2d & 1/3 & 2/3 & 0.25272(33) & 0.166(1) \\
& Nb(3) & 2d & 1/3 & 2/3 & 0.87707(33) & 0.166(1) \\
& O(1) & 6i & 0.16937(49) & 0.30685(19) & 0.30685(19) & 1/2 \\
& O(2) & 6i & 0.16413(38) & 0.83577(38) & 0.56995(19) & 1/2 \\
& O(3) & 6i & 0.17068(57) & 0.82922(57) & 0.93424(26) & 1/2 \\
& O(4) & 6i & 0.49774(61) & 0.50216(61) & 0.18738(14) & 1/2 \\
\multicolumn{1}{l}{} & \multicolumn{1}{l}{} & \multicolumn{1}{l}{} & \multicolumn{1}{l}{} & \multicolumn{1}{l}{} & \multicolumn{1}{l}{} & \multicolumn{1}{l}{} \\
 & \multicolumn{6}{c}{$a$ = 5.7902(2) ({\AA}),  $c$ = 18.9026(3) ({\AA})} \\
\multicolumn{1}{l}{} & \multicolumn{6}{l}{} \\
\multicolumn{1}{l}{} & \multicolumn{6}{c}{Overall B-factor = 0.093(8) ({\AA}$^{2}$)} \\ \hline
\end{tabular}

\end{table*}

\textit{Magnetic susceptibility scaling---}%
The temperature dependence of the AC susceptibility measured with a small AC field and low frequency should reflect the intrinsic susceptibility behavior of a system, or has the same temperature trend of the DC susceptibility measured on the same system. Therefore, a  AC field of 0.5 Oe with frequency 300 Hz was used to measure the magnetic susceptibility down to 0.3 K. This data was easily matched to the high temperature DC susceptibility data taken down to 1.8 K with a simple scaling factor (Fig. S1(b)). As the behaviors between them are essentially identical, the scaled AC data was used for all simulations and fittings to extend the data down to 0.3 K. Unit of the DC susceptibility $\chi$ = emu/mol is used for scaled AC data to maintain continuity. \\

\begin{figure}[tbp]
	\linespread{1}
	\par
	\begin{center}
		\includegraphics[width= 6 in]{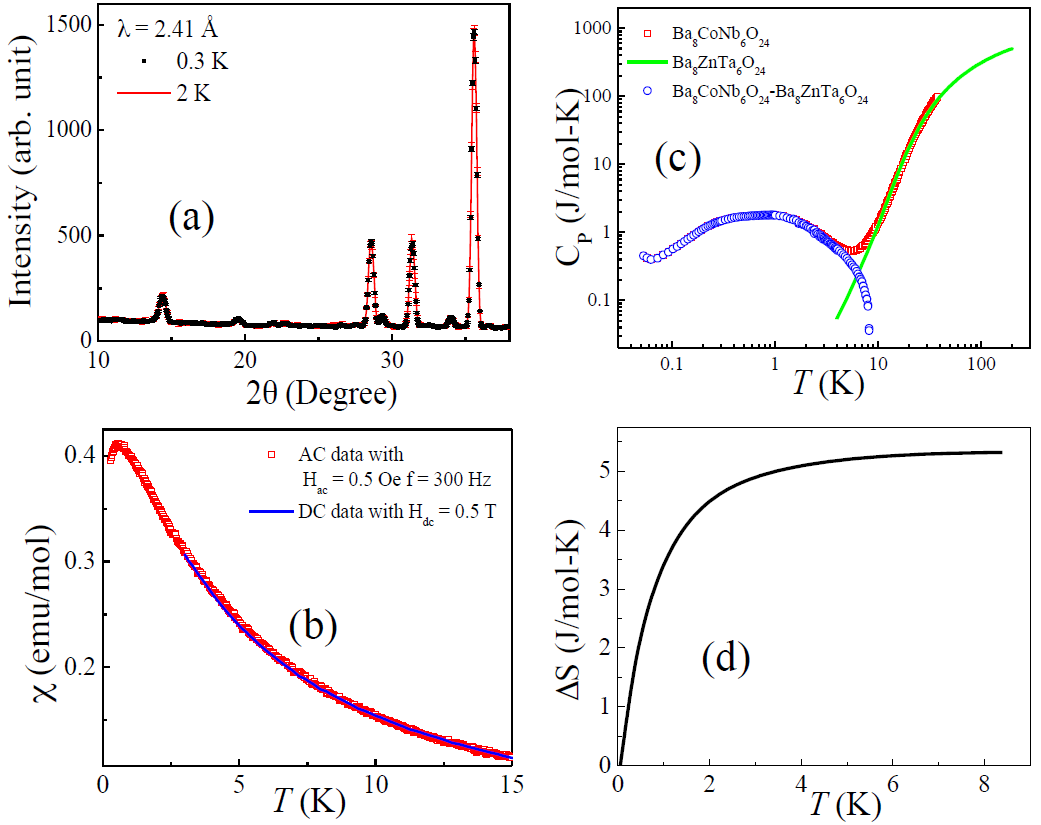}
	\end{center}
	\par
	\caption{\label{diff}%
      (color online)
      (a) The neutron powder diffraction patterns for Ba$_8$CoNb$_6$O$_{24}$ taken at 0.3 and 2 K at HB-2A with  $\lambda$ =  2.4111 {\AA}. (b) The magnetic susceptibility for Ba$_8$CoNb$_6$O$_{24}$ shown with DC data and scaled AC data. (c) The specific heat for Ba$_8$CoNb$_6$O$_{24}$, Ba$_8$ZnTa$_6$O$_{24}$ and the difference taken to isolate the magnetic contribution. (d) The magnetic entropy calculated from integration of $C_{\text{P}}$/$T$. Error bars correspond to uncertainties of one standard deviation.}
\end{figure}

\textit{Specific heat---}%
The magnetic specific heat for Ba$_8$CoNb$_6$O$_{24}$ was isolated by subtracting a non-magnetic lattice standard, Ba$_8$ZnTa$_6$O$_{24}$ (Fig. S1(c)). Since the high temperature series expansion (HTSE) only provides magnetic contributions to the $C_{\text{P}}$, the standard subtracted data was used in all fittings and comparisons. The integration of $C_{\text{P}}$/$T$ between 0.06 and 8 K of the magnetic $C_{\text{m}}$ saturates at $\Delta$S = 5.32 J/mol-K. This value is close to the Rln(2)= 5.76 J/mol-K, which is expected for a spin-1/2 system. \\

\textit{High temperature series expansion---}%
The HTSE was used to obtain the  exchange constant and exchange anisotropy of Ba$_8$CoNb$_6$O$_{24}$ in both $\chi$ and $C_{\text{P}}$ down to 0.5 K. The coefficients a$_n$, c$_n$ (Eqs. S1 and S2 respectively) for $C_{\text{P}}$ and $\chi$ of the spin-1/2 isotropic Heisenberg TLAF have been calculated up to 12th order in previous work~\cite{2a}. To analyze the effect of the exchange anisotropy, we computed the coefficients for the spin-1/2 {\it anisotropic} Heisenberg TLAF for up to 12th order  for $\Delta$ = 0.9, 1.1 (see Table~S2). $C_{\text{P}}$ was calculated from the $T$-derivative of the free energy given in Eq. S1 and $\chi$ was calculated using Eq. S2. \\

\begin{equation}
\displaystyle{
\label{eq:B}
\frac{{\text {ln}} Z}{N}=\sum_{n=0}a_nx^{^n}}
\end{equation}

\begin{equation}
\displaystyle{
\label{eq:C}
\frac{k_BT{\chi}}{(g{\mu}_B)^2}=\sum_{n=0}c_nx^{^n}
}
\end{equation}

\textit{Pad\'e approximants---}%
The low-temperature divergence of the plain HTSE  of Eq.~S1 can be avoided  by using  the ratio method, integrated differential approximants, and Pad\'e approximants \cite{2g}. For this letter, we use Pad\'e approximants of order [m = 6, n = 6] as originally used by Elstner et al \cite{2a} and described in Eq.~S3. We chose approximants of order [6, 6]  for two reasons. First, $C_{\text{P}}$ being calculated by differentiating free energy results in non-physical divergences at $T$ = $J$/k$_{\text{B}}$ for some off-diagonal approximants, namely [7, 5] and [5,  7]. Second,  the diagonal order [n, n]  avoids a divergent low-temperature behavior. Other approximants were examined, but Pad\'e approximants of order [6,6] provide the most physically meaningful results for simulations below $T$ = $J/k_{\text{B}}$.
\begin{equation}
\displaystyle{
\label{eq:A}
R(x)=\frac{\sum_{j=0}^ma_jx^j}{1+\sum_{k=1}^na_kx^k}
}
\end{equation}

\begin{table*}[tbp]
\par
\caption{HTSE coefficients for $\Delta$ = 0.9, 1.1}
\label{t2}
\setlength{\tabcolsep}{0.25cm}
\begin{tabular}{ c c c c c}
\hline
\hline
\multicolumn{1}{l}{n} & \multicolumn{2}{l}{$\Delta = 0.9$} & \multicolumn{2}{l}{$\Delta = 1.1$}\\
&  ln$Z$ & $\chi$ & ln$Z$ & $\chi$ \\
0 & 0.693147180560E+00 & 0.250000000000E+00 & 0.693147180560E+00 & 0.250000000000E+00\\
1 & 0.000000000000E+00 & -0.135000000000E+01 & 0.000000000000E+00 & -0.165000000000E+01\\
2 & 0.421500000000E+01 & 0.457500000000E+01 & 0.481500000000E+01 & 0.757500000000E+01\\
3 & -0.275800000000E+01 & -0.928600000000E+01 & -0.336200000000E+01 & -0.267540000000E+02\\
4 & -0.110157250000E+02 & 0.735865000000E+01 & -0.143937250000E+02 & 0.833306500000E+02\\
5 & 0.227150400000E+02 & 0.813502200000E+01 & 0.316609600000E+02 & -0.274087222000E+03\\
6 & 0.555289564000E+02 & -0.432566133333E+01 & 0.830983377333E+02 & 0.971839778667E+03\\
7 & -0.230309437387E+03 & 0.309929443029E+02 & -0.367686192480E+03 & -0.315219402327E+04\\
8 & -0.251495874650E+03 & -0.905870806367E+03 & -0.428893232275E+03 & 0.868118706817E+04\\
9 & 0.237935023053E+04 & 0.359542189692E+04 & 0.433229044400E+04 & -0.243590877456E+05\\
10 & -0.490854208087E+02 & -0.374861931449E+02 & -0.625397496451E+02 & 0.877627224099E+05\\
11 & -0.239243887884E+05 & -0.309077579057E+05 & -0.498804497089E+05 & -0.317811658164E+06\\
12 & 0.262699694109E+05 & 0.180327243359E+05 & 0.582734814011E+05 & 0.840953315702E+06\\
\hline
\end{tabular}

\end{table*}

\textit{Inelastic neutron scattering---}%
Fig. S2 shows the INS spectra measured from 0.06 and 5 K. There is no sign for LRO or structural distortion down to 0.06 K. The magnetic dynamic signal begins to appear around  $T$ = 5 K. With decreasing temperature, the magnetic spectra becomes more clear. However, the INS spectra does not change below $T$ = 0.5 K and the spin-wave is not well formed even at 0.06 K. \\

\textit{Spin-wave theory---}%
The isotropic Hamiltonian with only nearest neighbor exchange interaction $J = 0.144$ meV was considered for both linear spin-wave theory and spin-wave theory with $1/S$ correction. The linear spin-wave theory simulation was performed using SpinW \cite{spinw} program. The powder averaged spectrum was calculated using simple Monte Carlo integration with sampling size for each $Q$-$E$ point set to $10^5$. Spin-wave theory with $1/S$ correction was calculated based on the results derived by Mourigal \textit{et al.} \cite{tafdsf}. The numerical integrations including powder averaging were implemented using the adaptive multidimensional integration (cubature) algorithm with the maximum number of integrated points set to $2\times10^3$. Data in both cases were convoluted with estimated instrumental energy resolution $\Delta E = 0.025$ meV and $Q$ resolution $\Delta Q = 0.015$ \AA$^{-1}$ which was approximated by fitting the nuclear peak width. In order to make easier comparison between the simulated results and experimental data, we reduced the former into the same $Q$-$E$ grid as the latter using spline interpolation. \\

\begin{figure}[tbph!]
	\linespread{1}
	\par
	\begin{center}
		\includegraphics[width= 6 in]{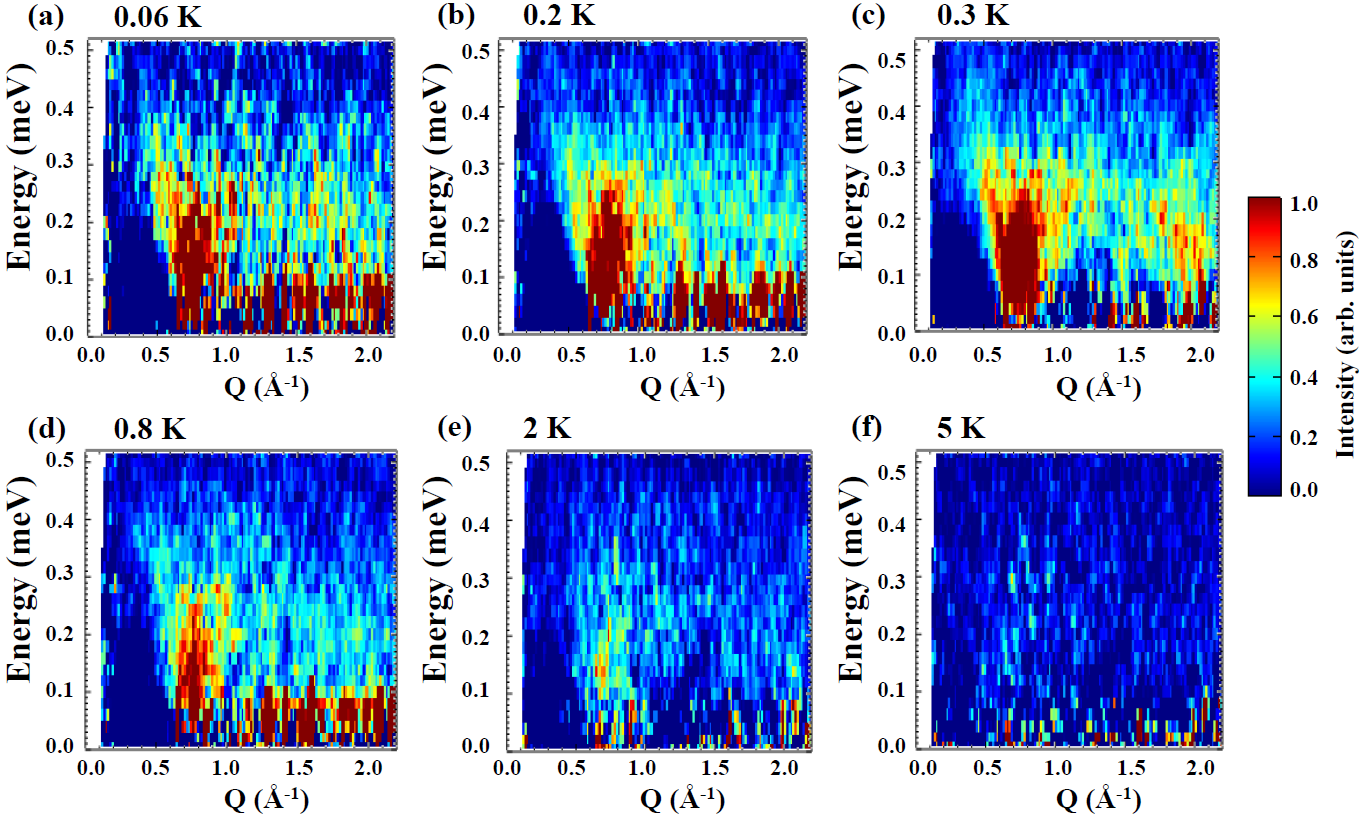}
	\end{center}
	\par
	\caption{\label{diff}%
      (color online)
       The temperature dependence of INS spectra for Ba$_8$CoNb$_6$O$_{24}$. The back ground had been subtracted.}
\end{figure}


\begin{thebibliography}{99}

\bibitem{Mermin_1966} N. D. Mermin and H. Wagner, Phys. Rev. Lett. {\bf 17}, 1133 (1966).

\bibitem{Lieb_1961} E. Lieb, T. Schultz, and D. Mattis, Ann. Phys. {\bf 16}, 407 (1961).

\bibitem{Faddeev_1981} L. D. Faddeev and L. A. Takhtajan, Phys. Lett. A {\bf 85}, 375 (1981).

\bibitem{Tennant_1993} D. A. Tennant, T. G. Perring, R. A. Cowley, and S. E. Nagler, Phys. Rev. Lett. {\bf 70}, 4003 (1993).

\bibitem{Mourigal_2013} M. Mourigal, M. Enderle, A. Kl{\"o}pperpieper, J.-S. Caux, A. Stunault, and H. M. R\o{}nnow, Nature Physics {\bf 9}, 435 (2013).

\bibitem{Savary_2016}  L. Savary and L. Balents, arXiv:1601.03742 (2016).

\bibitem{Banerjee_2016} A. Banerjee, C. A. Bridges, J.-Q. Yan, A. A. Aczel, L. Li, M. B. Stone, G. E. Granroth, M. D. Lumsden,	Y. Yiu, J. Knolle, S. Bhattacharjee, D. L. Kovrizhin, R. Moessner, D. A. Tennant, D. G. Mandrus, and S. E. Nagler, Nature Materials, doi:10.1038/nmat4604 (2016).

\bibitem{Jolicoeur89} %Analytical
Th. Jolicoeur and J. C. Le Guillou, Phys. Rev. B {\bf 40}, 2727 (1989).

\bibitem{Chubukov94} %Analytical
A. V. Chubukov, S. Sachdev, and T. Senthil, J. Phys: Cond. Matt. {\bf 6}, 8891 (1994).

\bibitem{Capriotti99} %QMC
L. Capriotti, A. E. Trumper, and S. Sorella, Phys. Rev. Lett. {\bf 82}, 3899 (1999).

\bibitem{Zheng06} %SE
W. H. Zheng,  J. O. Fj\ae restad, R. R. P. Singh, R. H. McKenzie,and R. Coldea, Phys. Rev. B {\bf 74}, 224420 (2006).

\bibitem{White06} %DMRG
S. R. White and A. L. Chernyshev, Phys. Rev. Lett. {\bf 99}, 127004 (2007).

\bibitem{Chernyshev_2009} A. L. Chernyshev and M. E. Zhitomirsky, Phys. Rev. B. {\bf 79}, 144416 (2009).

\bibitem{r} K Hirakawa, J. Appl. Phys. {\bf 53}, 1893 (1982).
\bibitem{w} A. Cuccoli, T. Roscilde, R. Vaia, and P. Verrucchi, Phys. Rev. Lett. {\bf 90}, 167205 (2003).
\bibitem{b} S. Miyashita and H. Kawamura, J. Phys. Soc. Jpn. {\bf 54}, 3385 (1985).
\bibitem{m} W. Stephan and B. W. Southern, Phys. Rev. B. {\bf 61}, 11514 (2000).
\bibitem{o} S. Fujimoto, Phys. Rev. B. {\bf 73}, 184401 (2006).

\bibitem{Coldea_1997} R. Coldea, D. A. Tennant, R. A. Cowley, D. F. McMorrow, B. Dorner, and Z. Tylczynski, Phys. Rev. Lett. {\bf 79}, 151 (1997).

\bibitem{2d} Y. Doi, Y. Hinatsu, and K. Ohoyama, J. Phys: Cond. Mat. {\bf 16}, 8923 (2004).

\bibitem{2h} H. Tsujii, C. R. Rotundu, T. Ono, H. Tanaka, B. Andraka, K. Ingersent, and Y. Takano, Phys. Rev. B. {\bf 76}, 060406 (2007).
\bibitem{1i} W.-J. Hu, S.-S. Gong, W. Zhu, and D. N. Sheng, Phys. Rev. B. {\bf 92}, 140403 (2015).
\bibitem{Koutroulakis2015} G. Koutroulakis, T. Zhou, Y. Kamiya, J. D. Thompson, H. D. Zhou, C. D Batista, and S. E. Brown, Phys. Rev. B. {\bf 91}, 024410 (2015).

\bibitem{2m} J. Ma, Y. Kamiya, T. Hong, H. B. Cao, G. Ehlers, W. Tian, C. D. Batista, Z. L. Dun, H. D. Zhou, and M. Matsuda, Phys. Rev. Lett. {\bf 116}, 087201 (2016).
\bibitem{2c} Y. Shirata, H. Tanaka, A. Matsuo, and K. Kindo, Phys. Rev. Lett. {\bf 108}, 057205 (2012).
\bibitem{2e} T. Susuki, N. Kurita, T. Tanaka, H. Nojiri, A. Matsuo, K. Kindo, and H. Tanaka, Phys. Rev. Lett. {\bf 110}, 267201 (2013).

\bibitem{2i} N. A. Fortune, S. T. Hannahs, Y. Yoshida, T. E. Sherline, T. Ono, H. Tanaka, and Y. Takano, Phys. Rev. Lett. {\bf 102}, 257201 (2009).
\bibitem{2o} H. D. Zhou, C. Xu, A. M. Hallas, H. J. Silverstein, C. R. Wiebe, I. Umegaki, J. Q. Yan, T. P. Murphy, J. -H. Park, Y. Qiu, J. R. D. Copley, J. S. Gardner, and Y. Takano, Phys. Rev. Lett. {\bf 109}, 267206 (2012).

\bibitem{Ghioldi_2015} E. A. Ghioldi, A. Mezio, L. O. Manuel, R. R. P. Singh, J. Oitmaa, and A. E. Trumper, Phys. Rev. B {\bf 91}, 134423 (2015).

\bibitem{2p} P. M. Mallinson, M. M. Allix, J. B. Claridge, R. M. Ibberson, D. M. Iddles, T. Price, and M. J. Rosseinsky, Angew. Chem Int. Ed. {\bf 44}, 7733 (2005).
\bibitem{supp} See online supplementary information.
\bibitem{ACoB3} M.~F.~Collins and O.~A.~Petrenko Can. J. Phys. {\bf 75}, 605 (1997).
\bibitem{2g} P. Sengupta, A. W. Sandvik, R. R. P. Singh, Phys. Rev. B. {\bf 68}, 094423 (2003).
\bibitem{2q} M. Lee, J. Hwang, E. S. Choi, J. Ma, C. R. Dela Cruz, M. Zhu, X. Ke, Z. L. Dun, and H. D. Zhou, Phys. Rev. B. {\bf 89}, 104420 (2014).
\bibitem{2t} N. Chandrasekharan and S. Vasudevan, Phys. Rev. B. {\bf 54}, 14903 (1996).
\bibitem{htse1} H. Rosner, R. R. P. Singh, W. H. Zheng, J. Oitmaa, and W. E. Pickett, Phys. Rev. B. {\bf 67}, 014416 (2003).
\bibitem{htse2} R. R. P. Singh and J. Oitmaa, Phys. Rev. B. {\bf 85}, 104406 (2012).
\bibitem{htse3} J. Oitmaa, C. Hamer, and W. Zheng, Series Expansion Methods for Strongly Interacting Lattice Models, (University Press, Cambridge, 2006).
\bibitem{2a} N. Elstner, R. R. P. Singh, and A. P. Young, Phys. Rev. Lett. {\bf 71}, 1629 (1993).

\bibitem{tafdsf} M. Mourigal, W. T. Fuhrman, A. L. Chernyshev, and M. E. Zhitomirsky,Physical Review B {\bf 88}, 094407 (2013).
\bibitem{spinw} S. Toth and B. Lake, Journal of Physics: Condensed Matter {\bf 27}, 166002 (2015).
\bibitem{DallaPiazza} B. Dalla Piazza, M. Mourigal, N. B. Christensen, G. J. Nilsen, P. Tregenna-Piggott, T. G. Perring, M. Enderle, D. F. McMorrow, D. A. Ivanov, and H. M. R\o{}nnow, Nature Physics {\bf 11}, 62 (2015).

\bibitem{AC} Z. L. Dun, E. S. Choi, A. M. Hallas, C. R. Wiebe, J. S. Gardner, E. Arrighi, R. S. Freitas, A. M. Arevalo-Lopez, J. P. Attfield, H. D. Zhou, and J. G. Cheng, Phys. Rev. B. {\bf 89}, 064401 (2014).
\bibitem{Fprof} J. Rodriguez-Carvajal, Physica B {\bf 192}, 55-69 (1993).
\bibitem{Copley} J. R. D. Copley and J. C. Cook, $Chem.$ $Phys.$ {\bf 292}, 477 (2003).
\bibitem{2a} N. Elstner, R. R. P. Singh, and A. P. Young, Phys. Rev. Lett. {\bf 71}, 1629 (1993).
\bibitem{2g} J. Oitmaa, C. Hamer, and W. Zheng, Series Expansion Methods for Strongly Interacting Lattice Models, (University Press, Cambridge, 2006).
\bibitem{spinw} S. Toth and B. Lake, Journal of Physics: Condensed Matter {\bf 27}, 166002 (2015).

\end{thebibliography}
\end{document}